# Geometric Phase in Optics and Angular Momentum of Light

*S.C. Tiwari*
*Institute of Natural Philosophy*
*1, Kusum Kutir, Mahamanapuri*
*Varanasi-221 005, India*

## Abstract

Physical mechanism for the geometric phase in terms of angular momentum exchange is elucidated. It is argued that geometric phase arising out of the cyclic changes in the transverse mode space of Gaussian light beams is a manifestation of the cycles in the momentum space of the light. Apparent non-conservation of orbital angular momentum in the spontaneous parametric down conversion for the classical light beams is proposed to be related with the geometric phase.

Phase singularities in light beams had been an active area of research, however prior to 1992 [1] their relationship with a well defined orbital angular momentum (OAM) of electromagnetic (EM) waves was not clear. Paraxial wave solutions of the source free Maxwell equations with helical wavefronts were shown to possess OAM in the units of ž (Planck constant/$2\pi$) in [1], and a nice early exposition of the phase singularities and their generation using computer-generated holograms can be found in [2]. For a recent review we refer to [3]. On the other hand, stimulated by the Berry phase in quantum mechanics [4], its optical analogues i.e. Pancharatnam phase in the polarization state space (geometrically represented by the Poincare sphere) and Rytov-Vladimirskii phase in the wave-vector space (geometrically represented by the sphere of directions in the momentum space) were rediscovered. There exists a considerable debate on the question whether GP in optics is a classical or a quantum phenomenon, see [5] and also the review [6], however the issue of the physical origin of GP was addressed in [5]. It was suggested that spin and orbital parts of the angular momentum were responsible for the Pancharatnam and Rytov-Vladimirskii-Chiao-Wu (RVCW) phases respectively. [Note that Chiao and Wu proposed the spin redirection phase [7] similar to the earlier work of Rytov-Vladimirskii, hence the name RVCW phase]. Earlier Jiao et al [8] considered both phases simultaneously, and used the geometry of a generalized Poincare sphere to suggest that the angular momentum exchange of light with optical elements was common origin of both the phases. Later the analyses by van Enk [9] and Banerjee [10] supported the hypothesis [5] that angular momentum exchange was a physical mechanism for the GP. In [9] van Enk proposed that geometric phase arising out of the cycles in the mode space of Gaussian light beams was a new phase. A Poincare-sphere equivalent for OAM states was proposed by Padgett and Courtial [11], and rotational frequency shift in a rotating mode converter for OAM bearing beams was interpreted in terms of this GP. In a recent paper, Galvez et al [12] claim to have made first direct measurement of GP in the mode space. The meaning of angular momentum exchange remains obscure in the cited works though we clearly related it with the angular momentum holonomy in [5], and gave a plausibility argument for the gauge theoretic approach for the Pancharatnam phase in [13].

In this paper we address three issues. Since the light beams with OAM are now well established, it is possible to elucidate the abstract proposition put forward in [5] in a physically transparent manner. Secondly the analysis of the physical origin of GP shows that the GP in mode space is a manifestation of RVCW phase. And lastly, we ask the question: is missing angular momentum in the spontaneous parametric down-conversion (SPDC) observed for the classical Laguerre-Gaussian (LG) beams [14] related with GP?

Mathematically, GP could be viewed as a parallel transport holonomy in the real or complex spaces with non-trivial geometries and topologies. Meaning of holonomy becomes clear if we take a simple example. Let us consider the parallel transport of a vector on the surface of the sphere around a closed loop, then the direction of the vector gets rotated i.e. the vector does not return to its original direction due to the curved geometry of the surface. Mathematicians term it holonomy though Berry argues that it would be more appropriate to call it anholonomy [5]. If the holonomy makes physical sense, then following Berry's analysis [4] one usually seeks a suitable parameter space, and distinguishes dynamical and geometrical effects. In [5, 13] we asked: Is there a fundamental physical origin for the geometrical effects? To put the problem in perspective, recall that the Aharonov-Bohm (AB) effect is usually believed to be a typical quantum effect in the literature since its original formulation in [15]. It was shown in [15] that there could be an observable effect of changing the EM field on the electron moving in a field free region if the problem was treated quantum mechanically. The AB effect has been demonstrated experimentally, and in spite of vast literature on this subject [16] the physical reality of the EM potentials remains debatable. AB effect appears counter-intuitive because in the classical Newton-Lorentz equation of motion for the electron it is only the EM fields that occur, and in the field free region there is no dynamics. In our work on the foundations of relativity [17], we have arrived at the conclusion that the inertial frames are non-equivalent, and the effect of unobservable uniform potentials that characterize the inertial frames could manifest in certain circumstances. The dynamics described by the force (torque) law in terms of the rate of change of momentum (angular momentum) is inadequate to account for such effects. It is no surprise then that the AB effect and Berry phase first appeared in quantum mechanics since the Schrodinger equation is based on the Hamilton-Jacobi equation in which momentum is derived from a scalar Hamilton's principal function. Similarly, a classical analogue of Berry phase is found in a variant of the Hamilton-Jacobi formulation i.e. the action-angle variables [18]. In a sense a pre-dynamical state can be envisaged in these formulations. Thus our proposition is that the geometrical or topological effects have their physical origin in the pre-dynamics.

To the specific case of GP in optics, in [5] an attempt was made to trace its origin to the angular momentum holonomy arising from the pure gauge potentials. The spin (orbital) part of the angular momentum exchange is related with the Pancharatnam (RVCW) phase of light. In the light of the preceding discussion, though torque may vanish, the constant level of angular momentum may change after the completion of a cycle on the Poincare sphere (or wave vector, **k**-space) and manifest as a geometrical effect. For example, if the light beam prepared in a definite polarization state is made to traverse intermediate optical elements (e.g. wave plates) then there is a transfer of spin angular momentum at each transformation stage of the polarization, and though this is necessary for the Pancharatnam phase it is not sufficient. The polarization cycle must include a non-zero solid angle on the Poincare sphere i.e. if the polarization changes are such that the path is re-traced, then the GP is zero. The spin angular momentum exchange responsible for the Pancharatnam phase is the net shift in the spin angular momentum, ΔS of the beam after the completion of the cycle. It is known that the spin per photon is $\pm \check{z}$, therefore though the polarization of the beam is restored there must occur changes in the photon numbers as pointed out in [13]. It may be cautioned that in quantum optics, total number of photons is proportional to the square of the electric field amplitude, however one has to be careful calculating the photon density in the radiation field [19]. The net shift ΔS is the spin angular momentum holonomy, and depends on the photon number density or the electric field amplitude for the Pancharatnam phase.

In the spin redirection phase, the momentum space is considered for the cycles in the evolution of the light beam. In a typical case, a linearly polarized EM wave propagating along a helically wound optical fiber undergoes changes in the direction of the wave vector, and completes the circuit in the **k**-space (here momentum $=\check{z}\,\mathbf{k}$). The excess phase (RVCW phase) is equal to the solid angle subtended in the **k**-space. Since the EM field momentum is determined by the Poynting vector, **E** x **B** and the orbital angular momentum density is proportional to **r** x (**E** x **B**), we anticipate the origin of the RVCW phase in terms of the OAM exchange. Similar to the spin exchange case, here the OAM exchange with the optical elements during the traversal of the cyclic path in the **k**-space is necessary, and the beam acquires net shift in its OAM though the OAM per photon is restored to its initial value. This mechanism is not obvious in the Chiao-Wu scheme using optical fiber and plane wave propagation, however the OAM carrying light beams make it very transparent.

The light beams with spherical wavefronts do not possess OAM, while helical wavefronts with phase singularity on the beam axis [2] are shown to bear OAM in the propagation direction z in the cylindrical coordinate system (r, $\phi$, z) [1]. The $\phi$-component of the Poynting vector gives rise to the azimuthal linear momentum that leads to OAM. The Poynting vector spirals along the z-direction for such a case. Note that a Hermite-Gaussian (HG) beam with its axis along z-direction has zero OAM. The field amplitude of $HG_{nm}$ mode is a product of Hermite polynomials of order n and m, and a Gaussian function. The field amplitude for the $LG_p^l$ mode is proportional to the product of a Gaussian function with the associated Laguerre polynomial $L_p^l$. HG and LG modes are solutions of the scalar Helmholtz equation in paraxial approximation in rectangular and cylindrical coordinate system respectively. Azimuthal index $l$ is the number of $2\pi$ cycles in phase around the circumference, and p+1 is the number of radial nodes for the $LG_p^l$ mode. The interpretation that this beam carries OAM of $\hbar$ per photon makes sense, if and only if, the Poynting vector and angular momentum density of the EM field are calculated, and the ratio of the flux of angular momentum to the EM energy is obtained to be $l/\omega$ ($=\hbar l/\hbar \omega$). For the scalar field amplitude of LG mode, OAM has no meaning. Mathematically both HG and LG modes form complete basis sets, and one can be represented by a linear superposition of another set. The transformation $HG_{mn} \rightarrow LG_p^l$ is called the mode conversion. Here $l$ = m-n and p = min (m, n). For details see the review [3].

Any physical realization of the mode conversion involves changes in the wave vector of the beam: distribution of intensity lobes in rectangular symmetry, planar wavefronts, and Poynting vector directed along the direction of propagation are characteristic features of the HG modes while circular cross-section and zero intensity along the beam axis, helical wavefronts, and spiraling Poynting vector characterize the LG modes. There are several techniques to achieve the mode conversion: holography, spiral phase-plates, cylindrical lenses and optical fibers. To illustrate the significance of linear momentum or wave vector changes in the process of mode conversion let us consider the spiral phase-plates. These are transparent glass discs having thickness increasing with azimuthal angle giving rise to a radial step in the surface. The height of the step corresponds to an integral multiple of wavelength of the incident light for a specific refractive index of the glass. Incident HG beam after passing through the phase-plate acquires a helical phase structure. Simple analysis shows that after refraction the beam acquires an azimuthal linear momentum, and hence an OAM. In the cylindrical lens system $\pi/2$ and $\pi$ mode converters act like quarter-wave plate and half-wave plate birefringent optical elements used for polarization changes respectively for the OAM transformations. For an incident HG mode on a cylindrical lens the Gouy phase shift depends on the mode indices and the astigmatism of the lens. Rotated HG mode and intermediate elliptical beam waist imply changes in the Poynting vector or wave number.

Now it becomes clear that it is not merely the mode space, but the underlying Poynting vector transformations that have physical significance for the optical properties of the beams. Since the linear momentum (determined by Poynting vector) cycles in the **k**-space give rise to the geometrical phase, for the Gaussian beams also there would exist an analogue of the RVCW phase. It is reasonable to argue that the recently observed GP in mode transformations [12] is a manifestation of the RVCW phase since transformation of $LG_0^{-1}$ beam are achieved by wave vector transformations. Analogies have heuristic values and may provide new insights, however the construction of orbital Poincare sphere [11] using mode functions may be misleading. Unlike spin angular momentum having only two values, OAM does not have any such restriction. Polarization of classical light beams is an intrinsic property, while for OAM bearing beams the Poynting vector plays the crucial role and as argued earlier scalar field amplitude modes by themselves are not sufficient to exhibit OAM – a fact demonstrated by the work of Allen et al [1]. For the same reason, the proposition of van Enk [9] that GP in the mode space of the Gaussian beams is a new effect becomes an artifact; instead we argue that any cyclic mode transformation that traverses a closed path in the **k**-space would acquire a GP, therefore calculation of Poynting vector for intermediate states during the traversal of the path is essential for establishing the existence of the GP for the Gaussian beams.

It is also pertinent to note that this interpretation remains valid if a Dove prism instead of lens system is used to reverse the handedness of the mode. In this case we have to use the discrete transformations of perfect mirror reflections discussed by Kitano et al [20]. Equivalence of sudden change in the wave vector upon reflection, and adiabatic traversal of a path in a modified **k-**space shown in [20] implies that the mode conversion using a Dove prism (mirror-like optical element) could be represented in terms of continuous changes achieved by cylindrical lenses. Of course, a detailed analysis would be necessary.

Amongst several studies on the OAM bearing light beams, parametric down-conversion has turned out to be controversial in its theoretical interpretation. The process of SPDC involves interaction of a pump field with a nonlinear crystal to generate signal and idler fields. Conservation of energy and phase-matching conditions must be satisfied in this process: first condition for a degenerate SPDC means that both idler and signal beams have frequency exactly half of the pump frequency for a lossless crystal, while phase-matching is a somewhat subtle issue. Reported observations in [14] indicate non-conservation of OAM within the light beams, and this aspect is of interest in the present context. Let us first summarize the main experimental observations following [14]: i) classical LG pump beam generated by holograms with p=0 was passed through lithium triborate crystal cut for type-I noncritical phase matching, ii) the intensity profile of the down-converted beam was recorded by a classical set-up, iii) the variations in the intensity profiles as a function of distance behind the crystal were investigated for both pump and down-converted light, iv) for pump beam with $l=2$, the expected stable profile with zero on-axis intensity was found at various positions. On the other hand, down-converted beam was found to change its form, and at about 40 mm behind the crystal, the on-axis intensity minimum became unrecognizable, v) the same qualitative behaviour was observed when pump beam with $l=1$ was used, and vi) interference pattern of the down-converted beam with its sheared image established spatial incoherence of the down-converted beams. Later experiments performed at a single photon level confirmed the conservation of OAM [21]. Introducing $l$-entanglement and phase matching at a single photon level obscures the real problem arising out of the observations by Arlt et al [14] for the following reasons. Entanglement and EPR pairing are believed to be typical quantum characteristics, and there are fundamental questions related with them, see [22]. Physical nature of single photon, whether it is merely a calculation tool or has physical reality, is another intriguing question [22,23]. Subsequently apparent non-conservation of OAM [14] has been attributed to classical detection method [21], and Franke-Arnold et al [24] have made theoretical calculation of correlations between SPDC generated photon pairs using phase-matching at a single photon level to show OAM conservation. Note that $l$-entanglement itself embodies $l$-conservation, and the extrapolation of single photon result to classical light beams is not obvious. Arnaut and Barbosa [25] raise important points, namely the finite transverse size of the beam and symmetry of the nonlinear crystal, however the treatment lacks clarity, and they also employ single photon phase-matching in this process that does not appear convincing for the situation of the experiment in [14]. Comments and reply on this work [26] bring out the need for further study on the role of crystal symmetry and entanglement. Barbosa and Arnaut [27] clarify some aspects of their approach, in particular the phase-matching condition for finite-sized beams is elaborated, however the physical understanding of the apparent non-conservation of OAM for classical LG beams remains elusive.

We suggest a similar problem of apparent non-conservation of angular momentum first pointed out by Holbourn in 1936, and revived by Player in 1987 [28] may provide new insight to resolve this problem, and also may throw light on the angular momentum holonomy in geometric phases. Player related the missing angular momentum to the positional transverse shift of the light beam reflected from the interface between two lossless dielectric media. Hugrass [29] considers finite lateral and angular spreads of the beam, and calculates angular momentum from the first principles. The conservation law of angular momentum is established, however an interesting results is also obtained: if a coordinate system $(x', y', z')$ shifted by $(x_o, y_o)$ along x- and y-directions is used then the energy and momentum of the beam are the same as those in the original coordinate system $(x,y,z)$ but the angular momentum of the beam is changed by an amount equal to $(x_o p_y - y_o p_x)$. For specific cases, it corresponds to Player's transverse shift. In the analysis of Arnaut and Barbosa [25] the state vector for the down-converted light contains a function that depends on the azimuthal angles $\phi$ and $\phi'$. For the OAM balance, dependence of this function on $(\phi,\phi')$ is necessary excepting for the perfect collinearity. Comparing with the result of Hugrass, and the fact that the shift $(x,y) \to (x', y')$ is equivalent to $(\phi,\phi')$ it is tempting to conjecture that in both cases the missing angular momentum has same origin. Since the momentum of the beam in $(x,y,z)$ and $(x', y', z')$ remains unaltered, we can imagine a hypothetical **k**-space circuit, and associated GP due to the angular momentum shift. If the suggestion is correct, the idler and signal beams should originate with the phase difference equal to GP. It would be interesting to relate this phase difference with the low spatial coherence observed by Arlt et al [14], and the net OAM holonomy with the missing orbital angular momentum.

In this paper an attempt has been made to construct an integrated conceptual picture of the geometric phase in optics and the fundamental role of angular momentum. Tentative arguments at some places will have to be analyzed rigorously, and at least there are two open questions not touched upon in this article: (1) the phenomena of GP and angular momentum exchange at single photon level, and (2) the physical mechanism of rotational frequency shifts in various contexts, see the review [3] and also [30]. To conclude the paper, we present an outline of some proposed experimental schemes to test above ideas.

(1) If the **k**-space closed paths are basic to the GP in the mode space of the Gaussian beams then it would be useful to classify all HG and LG mode transformations that involve or do not involve the RVCW phases. In the examples that differ on the existence of GP according to our analysis and that of [9], the experiments of the kind carried out by Galvez et al [12] offer a feasible experimental test.

(2) For the plane EM waves, Pancharatnam and RVCW phases have been known to be additive [8]. For the Gaussian beams, a rigorous theory developed for nonparaxial case shows that spin and orbital parts of total angular momentum are not separable, and an additional correction term depending on spin is also present [3]. Recently Barnett shows that angular momentum flux is separable into spin and orbital parts in a gauge invariant form and without the paraxial approximation [31]. Barnett's conclusion is not surprising considering the fact that symmetric energy-momentum tensor is used by him. Starting with an action functional, and using the variational principle treating the electromagnetic potentials as independent field variables the natural Noether conserved quantity for infinitesimal coordinate transformations is the canonical energy-momentum tensor that is not gauge invariant. Third rank angular momentum tensor in this case has manifest presence of electromagnetic potentials. In fact we argued in [5] that the difference between gauge non-invariant and gauge invariant third rank tensors represented the holonomy responsible for GP. Note that angular momentum flux is incorporated in this formulation. Since Galvez et al [12] do not measure this holonomy their claim regarding angular momentum exchange is unfounded [32]. We suggest that rather than fixed polarization e.g. vertical linear polarization used in [12], it would be interesting to measure the GP of polarized Gaussian beam undergoing both polarization and **k**-space (or relevant mode space) cycles. This kind of experiment may give useful information on the physical nature of angular momentum tensor as well as on geometric phase.

(3) We have argued that GP is related with the angular momentum holonomy i.e. the shifts in the spin or OAM of the classical light beams accompanies the GP. This shift implies the changes in the levels of spin, and OAM off-axis of the beam. An interesting recent experiment by Garces-Chavez et al [33] claims to have measured spinning and orbital motion of a microscopic particle, and thus the local spin and OAM of the beam across the cross section. Radial dependence of spin and orbital rotation rates is found to be in agreement with the theory in their experiment. Authors note that absolute rotation rates are difficult to measure, therefore the shift in the levels of spin and OAM of the beams possessing GP cannot be directly measured. The schematic of this experiment will have to be slightly modified such that the motion of the particle is compared for the reference beam without GP, and the one with GP. If this experiment is technically feasible, then this could provide the direct evidence for the angular momentum holonomy and GP relationship.

**Acknowledgements:**

I am grateful to Prof. L. Allen for the encouragement and helpful correspondence. I thank Dr. S.J. van Enk Dr Dipti Banerjee, and Prof. E.J. Galvez for the references [29], [6] and [11] respectively. The Library facility at the Banaras Hindu University is acknowledged. I thank the reviewer for constructive comments.